\documentclass[%
 reprint,
 amsmath,
 amssymb,
 aps,
 prx,
 10pt,
 nofootinbib,
 floatfix,
 ]{revtex4-2}
\usepackage{pgf}
\usepackage{adjustbox}
\usepackage{listings}
\usepackage{graphicx}
\usepackage{dcolumn}
\usepackage{bm}
\usepackage[mathlines]{lineno}
\usepackage{tikz} 
\usetikzlibrary{quantikz}
\usepackage{xcolor}
\usepackage{etoolbox}
\usepackage{makecell}

\usepackage{hyperref}
\usepackage{cleveref}

\definecolor{rteal}{HTML}{00B5AD}
\definecolor{rmagenta}{HTML}{EF476F}
\hypersetup{urlcolor=rteal, citecolor=rmagenta, linkcolor=rmagenta, hyperindex, colorlinks, bookmarks=true}

\makeatletter
\patchcmd{\@verbatim}
  {\verbatim@font}
  {\verbatim@font\small}
  {}{}
\makeatother

\begin{document}

\preprint{APS/123-QED}
\title{Realization of arbitrary doubly-controlled quantum phase gates}
\author{Alexander D. Hill}\email[Corresponding author: ]{alex.hill@rigetti.com}
\affiliation{Rigetti Computing, 775 Heinz Avenue, Berkeley, CA, USA 94710}
\author{Mark J. Hodson}
\affiliation{Rigetti Computing, 775 Heinz Avenue, Berkeley, CA, USA 94710}
\author{Nicolas Didier}
\affiliation{Rigetti Computing, 775 Heinz Avenue, Berkeley, CA, USA 94710}
\author{Matthew J. Reagor}
\affiliation{Rigetti Computing, 775 Heinz Avenue, Berkeley, CA, USA 94710}

\date{\today}
\begin{abstract}
Developing quantum computers for real-world applications requires understanding theoretical sources of quantum advantage and applying those insights to design more powerful machines. Toward that end, we introduce a high-fidelity gate set inspired by a proposal for near-term quantum advantage in optimization problems. By orchestrating coherent, multi-level control over three transmon qutrits, we synthesize a family of deterministic, continuous-angle quantum phase gates acting in the natural three-qubit computational basis ($\mathtt{CCPHASE}(\theta)$). We estimate the process fidelity for this scheme via Cycle Benchmarking of $\mathcal{F}=87.1\pm0.8\%$, higher than reference two-qubit gate decompositions. $\mathtt{CCPHASE}(\theta)$ is anticipated to have broad experimental implications, and we report a blueprint demonstration for solving a class of binary constraint satisfaction problems whose construction is consistent with a path to quantum advantage.

\end{abstract}

\maketitle

\section{Introduction}
Quantum computers promise to perform tasks that are intractable with classical technologies---a feat known as quantum advantage~\cite{Feynman1982,Aaronson2011,Preskill2012,Bremner2016,Harrow2017,Bravyi2018,Boixo2018,Bouland2019}. Already, experiments sampling the output from pseudo-random configurations of superconducting~\cite{Arute2019,wu2021strong} and optical quantum computers~\cite{Zhong1460} have pushed beyond the reach of conventional high-performance computers. Building on the rigorous theoretical foundation of quantum advantage, and looking toward real-world applications, a remarkable connection has been established between a class of classically-hard sampling problems and a quantum approach to constraint satisfaction \cite{Bremner_2017,farhi2019quantum,Dalzell2020}, leading to the prediction of quantum advantage for Boolean satisfiability problems when encoded onto hundreds of qubits \cite{Dalzell2020}.

These problems concern the assignment of $n$ Boolean variables under a collection of constraint clauses ($C$). Each clause ($C_k$) is defined on at most three of the problem variables. The task is to determine solutions that maximize the number of constraints satisfied, known as MAX-3-SAT~\cite{Vazirani2010}. Applying the Quantum Approximate Optimization Algorithm (QAOA)~\cite{Farhi2014} quantum computers can, in principle, solve MAX-3-SAT by storing superpositions of solutions in registers of qubits~\cite{Wecker2016,Hadfield2019,Dalzell2020}. A key algorithmic step is then performing a diagonal phase-separation unitary associated to $C$: $U_C=\prod_k {U_{C_k}}$, with $U_k = \exp{\left(-i \gamma |x_k\rangle\langle x_k|\right)}$, where $x_k$ is a unique bitstring representation of $C_k$ and $\gamma$ is a variational parameter. Efficient large-scale emulations of this approach on classical hardware would imply the collapse of the polynomial hierarchy of complexity theories~\cite{Dalzell2020}. To date, however, the programmable $U_k$ phase gadget has been missing from the experimental toolbox.

Here, we experimentally demonstrate the requisite $U_k$ construction: a three-qubit $\mathtt{CCPHASE}$ gate, achieved with qutrit gate primitives based on the proposal in \cite{Abrams2020}, in turn extending the Toffoli gate technique from \cite{AFedorov2012}, that significantly reduces circuit depth over decompositions using only local, one- and two-qubit operations. We exploit the full digital phase control of parametrically-activated transmon-transmon interactions to produce a qutrit-based $\mathtt{CCPHASE}$ gate of an arbitrary phase. We benchmark such a gate on Aspen-9 -- a 32-qubit superconducting quantum processor. Finally, we discuss the successful application of $\mathtt{CCPHASE}$ gates to an algorithm of interest, QAOA for MAX-3-SAT (Section~\ref{section:qaoa}).

\section{Results}
\subsection{Multi-Level Coherent Control}
\label{section:pulse_description}

Hardware efficient schemes have been deployed to implement discrete versions of multiply-controlled gates (e.g., Toffoli), across a wide range of physical platforms \cite{Cory1998,Lanyon2009,Monz2009,AFedorov2012,Erhard2019,Levine2019}. Indeed, a number of important quantum computing tasks rely upon multiply-controlled gates, especially the Toffoli gate: reversible logic \cite{Vedral1996,Nielsen2011,Saeedi2013,Aaronson2015}, oracles for quantum search \cite{Grover1997,Figgatt2017} or simulation \cite{Low2019}, and quantum error correction \cite{Cory1998,Reed2012}. However, synthesizing these more complex unitaries using two-qubit gate sets typically results in prohibitive circuit depths. Recently, families of parameterized two-qubit entangling gates have been introduced, realizing improvements for specific near-term applications such as fermionic simulation \cite{Foxen2020} or combinatorial optimization \cite{Abrams2020}. Adapting these concepts for multiply-controlled gates unlocks the advantages of parameterized gates for key tasks. 

We first discuss the physics of our two-qubit $\mathtt{CPHASE}$ gates, from which we build the full three-qubit gate. Our current generation of superconducting quantum processors activate two-qubit entangling gates through AC modulation of magnetic flux applied to flux-tunable transmons \cite{Caldwell2018, NDidier2018, NDidier2019, Hong2020}. Each flux-tunable transmon qubit is statically coupled to a DC flux-tunable (but functionally fixed-frequency) neighbor. We can choose which qubit of the pair transitions to/from the second excited state in the $\mathtt{CPHASE}$ interaction (or, alternatively, activate an $\mathtt{iSWAP}$ interaction in the 0-1 subspace) by selecting the resonance condition (flux pulse frequency, amplitude, and duration) that places the correct modulation-induced sideband of the tunable transmon on resonance with its fixed-frequency neighbor. 

\begin{figure}
    \includegraphics[scale=0.9]{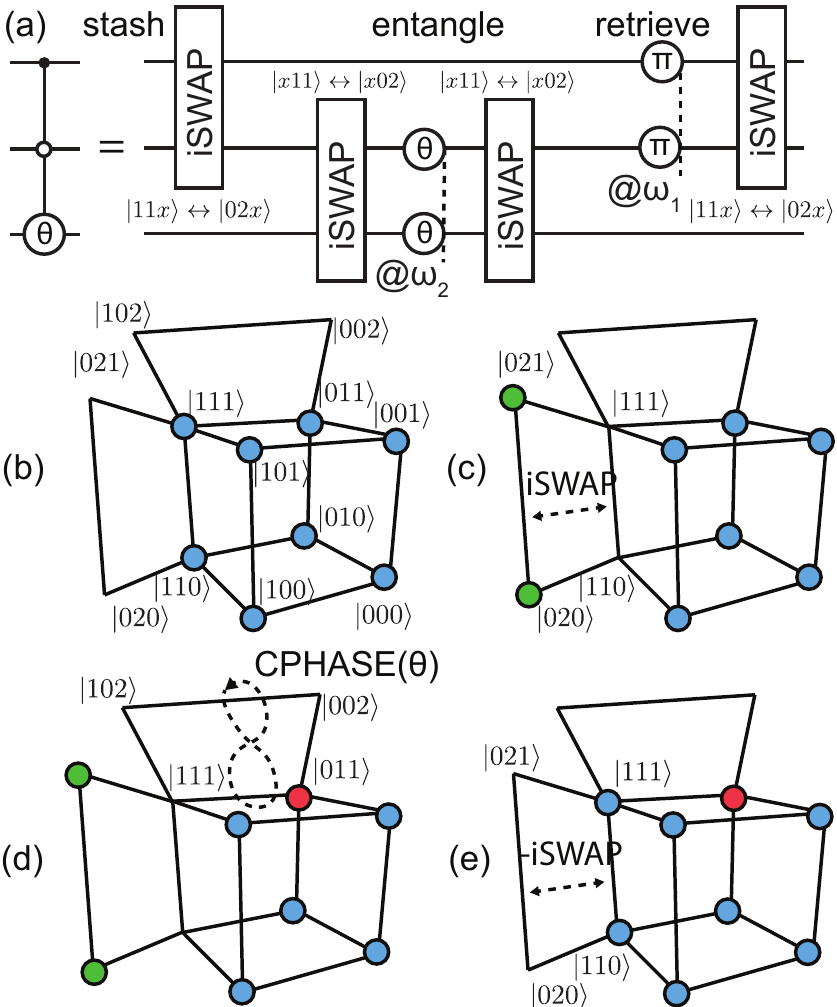}
    \caption{Method of operation of the three-qubit $\mathtt{CCPHASE}$ gate. (a) A circuit-level diagram of $\mathtt{CCPHASE}$ -- the circuit is equivalent to two arbitrary two-qubit $\mathtt{CPHASE}$ operations split and interdigitated to produce a multi-qutrit (SU(27)) circuit with a unitary in the standard computational basis (SU(8)). To apply the desired conditional phase only to the target state $|011\rangle$, we first store the competing states, $|11\textrm{x}\rangle$, by applying a single two-qutrit $\mathtt{iSWAP_{02/20}}$. A $\mathtt{CPHASE}$ operation on the second edge phases only the desired $|011\rangle$ state, as all others have been ``stashed,'' or are not phased by $\mathtt{CPHASE}$ in general. Finally, we retrieve the stored $|11\textrm{x}\rangle$ state using an $\mathtt{iSWAP}$ with a flux pulse phase calibrated to produce the identity for all states other than the target state of $|011\rangle$. Each step is mapped out in (b)-(e): (b) Subset of initial three-qutrit subspace. (c) Result of first $\mathtt{iSWAP}$ (``stash''). (d) Application of two-qubit conditional phase using a standard $\mathtt{CPHASE}$ on the second pair $|\mathrm{x}q_1q_2\rangle$. (e) A phase-shifted $\mathtt{iSWAP}$ operation returns the stashed states (``retrieval'') while unwinding any unwanted conditional phases on the first pair of qubits.
}
    \label{fig:overview_fig}
\end{figure}

Examining only the two-qubit interaction used to generate the $\mathtt{CPHASE}$ gate, the Hamiltonian takes the form:
\begin{align*}
\begin{split}
\hat{H}_{\textrm{int}} / \hbar &= \sum_{n=-\infty}^{\infty} g_n \left[ \sqrt{2}e^{i(n\omega_m - [\Delta_{02} - \eta_1])t}|02 \rangle\langle 11| \right. \\
                               &+ \left. \sqrt{2}e^{i(n\omega_m - [\Delta_{20} - \eta_0])t}|11 \rangle\langle 20| + \textrm{H.c.} \right] + ...,
\end{split}
\end{align*}
where $\omega_m$ is the frequency of the flux modulation on the flux-tunable qubit of the pair, $\eta_{0/1}$ are the anharmonicities of the first/second qubits, and $\Delta_{02/20}$ are the average detunings in the $|11\rangle-|02\rangle$ or $|11\rangle-|20\rangle$ subspaces, respectively. $g_n$ is the effective coupling of the interaction, which for small amplitudes is given by:
\[ g_n=g J_n (\epsilon/\omega_m )e^{-i\beta_n },\]
where $J_n$ are Bessel functions of the first kind and $\epsilon$ is the amplitude of the frequency oscillations. We directly control $\beta_n$ by changing the phase of the flux pulse that produces the interaction \cite{Reagor2018}.

We achieve arbitrary phase control of the $\mathtt{CPHASE}$ gate by splitting a complete $|11\rangle \leftrightarrow |02/20\rangle$ oscillation into two components. The individual flux pulses that produce the full $\mathtt{CPHASE}$ are equivalent to two-qu\textit{trit} gates - $\mathtt{iSWAP_{02}}$ or $\mathtt{iSWAP_{20}}$ -- representing $\mathtt{iSWAP}$ in the $|11\rangle-|02\rangle$ or $|11\rangle-|20\rangle$ (qutrit) subspace, respectively, with an off-diagonal controllable phase. Each AC flux-modulated qubit can produce both types of interaction: $\mathtt{iSWAP_{02}}$ corresponds to the transition $|11\rangle\rightarrow-i|02\rangle$, while $\mathtt{iSWAP_{20}}$ corresponds to the transition $|11\rangle\rightarrow -i|20\rangle$, for a given subsystem $|q_0 q_1\rangle$. The final conditional phase is a geometric phase produced by manipulating the phase relationship of the two pulses. Furthermore, by decomposing these $\mathtt{CPHASE}$ gates into their constituents, we obtain a set of flux-controllable qutrit operations that can be chained together to produce more sophisticated behavior in the standard qubit basis: for example, producing a controlled phase that is conditional on the state of more than one qubit.

We can derive the unitary resulting from such a sequence, shown in Figure~\ref{fig:overview_fig}, by considering the action of each qutrit operation in turn. The first $\mathtt{iSWAP_{02}}$ pulse acts on the first edge of the triplet, $|q_0q_1\rangle$, mapping $\{|00\mathrm{x}\rangle,|01\mathrm{x}\rangle,|10\mathrm{x}\rangle,|11\mathrm{x}\rangle\} \rightarrow \{|00\mathrm{x}\rangle,|01\mathrm{x}\rangle,|10\mathrm{x}\rangle,-i|02\mathrm{x}\rangle\}$. The second $\mathtt{iSWAP_{02}}$ pulse has the same effect on $|q_1q_2\rangle$, with transition $|x11\rangle\rightarrow-i|\mathrm{x}02\rangle$ occurring only when $|q_0 q_1\mathrm{x}\rangle=|01\mathrm{x}\rangle$, the only state where $q_1$ has remained in $|1\rangle$. This selects one element of the full SU(8), $|q_0 q_1 q_2\rangle=|011\rangle\rightarrow-i|002\rangle$, to be addressed with a controlled phase $\theta$ in the third $\mathtt{iSWAP_{02}}$ pulse. The phase of the physical flux pulse implementing the qutrit-qutrit interaction directly produces the desired conditional phase. The final $\mathtt{iSWAP_{02}}$ pulse returns the first edge to the computational subspace, i.e., maps $|q_0 q_1\rangle=-i|02\mathrm{x}\rangle\rightarrow|11\mathrm{x}\rangle$, where the additional phase $\pi$ is obtained from the flux phase, thereby producing the following unitary:
\begin{equation*}
\mathtt{CCPHASE}_{011}(\theta) = \text{diag}\left(1,1,1,e^{i\theta},1,1,1,1\right)
\end{equation*}
This qutrit-based $\mathtt{CCPHASE_{011}}$ gate is equivalent to the canonical $\mathtt{CCPHASE}$ gate (and others, e.g.,  $\mathtt{TOFFOLI}$) under single-qubit rotations. We note that this technique is analogous to that of Ref.~\cite{AFedorov2012}, with a few key advancements: first, we achieve arbitrary phase control of the doubly-controlled phase via manipulation of the underlying flux pulse; second, we reduce the overall interaction time by the equivalent of one full oscillation in the two-qubit space -- roughly 30\% of the overall gate time.

\begin{table*}
\centering
(a)
\begin{tabular}{c c c c c c c}
\hline
\hline
\makecell{Qubit \\ Index} & \makecell{$f_{01}$\\(GHz)} & \makecell{$f_{12}$\\(GHz)} & \makecell{$T_1^{(1)}$\\ $(\mu\textrm{s})$} & \makecell{$T_1^{(2)}$ \\$(\mu\textrm{s})$} & \makecell{$T_2^{01}$ \\ $(\mu\textrm{s})$} & \makecell{$T_2^{12}$\\$(\mu\textrm{s})$} \\
\hline
10 & 4.791 & 4.585 & 33.7(7) & 16.6(4) & 22(3) & 9.4(1) \\
11 & 3.247 & 3.044 & 46(8) & 18.9(8) & 24(9) & 1.6(3) \\
12 & 4.867 & 4.661 & 19(1) & 14.9(3) & 12.0(8) & 5.6(1) \\
\hline
\hline
\end{tabular}
\hspace{0.5cm}
(b)
\begin{tabular}{c c c c c}
\hline
\hline
Edge & \makecell{Interaction \\ Type} & \makecell{Pulse \\ Time \\ (ns)} & \makecell{Total $\mathtt{CPHASE}$ \\Gate Time \\(ns)} &  \makecell{$\mathtt{CPHASE}$ \\ Fidelity \\ (\%)}\\
\hline
(10,11) & $\mathtt{iSWAP_{20}}$ & 61 & 186 & 97.7(4)\\
(11,12) & $\mathtt{iSWAP_{02}}$ & 76 & 216 & 97.3(7) \\
\hline
\hline
\end{tabular}

\caption{Representative data for (a) three qubits on the Aspen-9 platform and (b) their associated gates. Qubits 10 and 12 are AC flux-tunable. The dephasing time with the second excited state is significantly reduced compared to the $0\leftrightarrow1$ dephasing time for qubit 11, which can impact overall gate performance if this qubit is held in the second excited state during our implementation of the three-qubit gate. The total CPHASE gate time consists of two pulses with 16 ns of rise/falltime and 16 ns of zero padding per pulse.}
\label{table:qubit_properties}
\end{table*}

\subsection{Experimental Verification}
One three-qubit sublattice from Aspen-9 was selected for experimental verification and benchmarking of the $\mathtt{CCPHASE_{011}}$ gate (see Table~\ref{table:qubit_properties} for qubit parameters and Ref.~\cite{Valery2021} for details regarding the design of the Aspen-9 quantum integrated circuit). For this site, we employed the interaction combination ($\mathtt{iSWAP_{20}}$, $\mathtt{iSWAP_{02}}$) which affects the intermediate transition $|011\rangle-i|002\rangle$, capable of producing a valid $\mathtt{CCPHASE_{011}}$. As discussed, our native $\mathtt{CPHASE}$ gates are constructed from two $\mathtt{iSWAP_{02/20}}$ pulses, so we may infer the fidelity of the individual $\mathtt{iSWAP_{02/20}}$ operations by taking the $\mathtt{CPHASE}$ fidelity as a strong lower bound. The edges (10, 11) and (11, 12) have associated two-qubit $\mathtt{CPHASE}$ fidelities of $97.7 \pm 0.4\%$ and $97.3 \pm 0.7\%$, respectively, estimated via interleaved Randomized Benchmarking~\cite{Magesan2012}. 

In general, designing the qutrit rotations such that only qubits with a large ratio of $E_J/E_C$ \cite{Koch2007} populate the second excited state are preferred. This can mitigate the effects of dephasing due to charge noise for qubits that enter the $|2\rangle$ state, and is especially important for $q_0$ (the first qubit of the triplet), which may stay in the $|2\rangle$ state while $q_1$ and $q_2$ are operated upon. On Aspen-9, these are the flux-tunable (even-numbered) qubits (on the Aspen architecture, those with the highest frequency); this correlates with the intermediate states $-i|200\rangle$ and $-i|002\rangle$ -- holding a $|2\rangle$ state component on qubits 10 or 12, respectively. 

 \begin{figure}
     \includegraphics[scale=0.9]{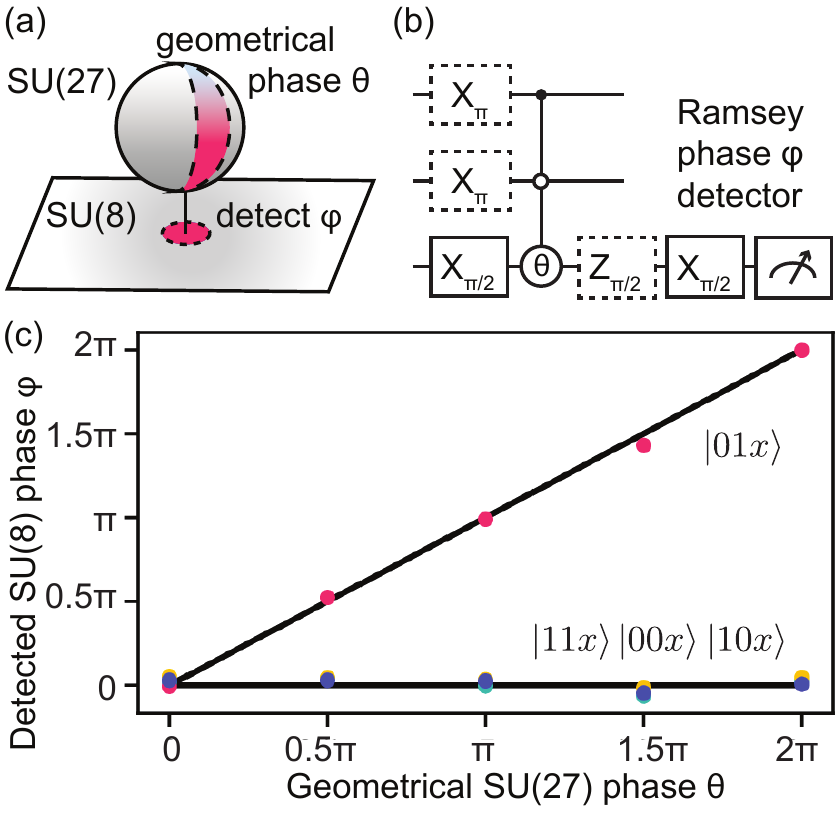}
     \caption{Verification of a calibrated $\mathtt{CCPHASE_{011}}$ gate. (a) We produce the target conditional phase in the three-qubit SU(8) space by manipulating two-qutrit gates existing in a higher-dimentional SU(27) space. The final conditional phase is equivalent to a geometric phase produced by controlling by the relative phases of the flux pulses producing each qutrit operation. (b) A Ramsey-interferometric experiment for determining the conditional phase applied as function of input state. Each phase is measured by selecting a target conditional phase, preparing an input control state, and measuring $\langle X \rangle$ and $\langle Y \rangle$ to determine the phase rotation accumulated by the target qubit on the Bloch sphere equator. (c) The  doubly-conditional phase is applied only when the control qubits are in the correct starting state $|01\rangle$, and the requested phase matches the recorded phase applied to the target qubit. }
     \label{fig:linear_phase_verification}
 \end{figure}
 
To verify the basic functioning of the calibrated $\mathtt{CCPHASE}$ gate, we examined a subset of the full three-qubit truth table and showed that the general features of the target unitary are reproduced accurately. This was done by performing a Ramsey-interferometric measurement of the phase evolution of the target qubit in the triplet after the gate was applied under a variety of input conditions -- in essence, a subset of full tomography for many target phases. As shown in \Cref{fig:linear_phase_verification}, we observe that the observed three-qubit conditional phase correlates linearly with the requested gate phase and that the conditional phase is only applied to the target qubit when the control qubits are in the correct input state $|01\textrm{x}\rangle$.

\begin{figure*}
    \centering
    \includegraphics[scale=0.3]{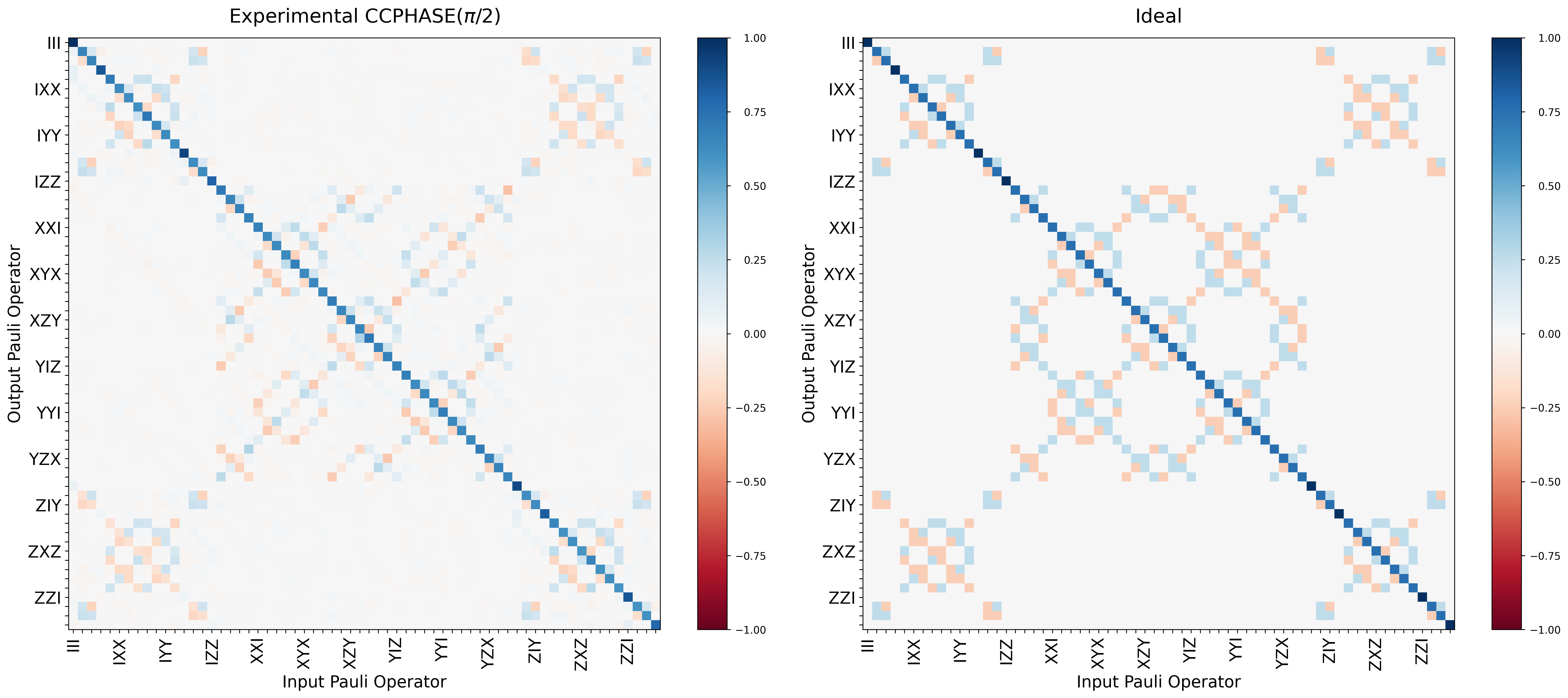}
    \caption{Quantum Process Tomography of $\mathtt{CCPHASE_{011}}$ for a target angle $\pi/2$ on Aspen-9 sublattice (10, 11, 12), with explicit calibration of $\mathtt{CPHASE}$ gate conditional phase on $|q_1 q_2\rangle$, yielding a fidelity of 81.8\%}
    \label{fig:qpt}
\end{figure*}

We applied three-qubit Quantum Process Tomography (QPT) \cite{Nielsen2011} as a verification that the gate produced the correct unitary for a specific conditional phase. QPT may be a very time- and resource-intensive process in the three-qubit case due to the exponentially increased number of measurements over the two-qubit case. Estimated fidelity is especially impacted by SPAM errors \cite{Merkel2013}, and can only provide a conservative lower bound. For these reasons, we perform QPT as a validation step for our calibration procedure, and do not repeat it for every benchmarked angle. The tomographic outputs and estimated process fidelities are reported in Figure~\ref{fig:qpt} at 81.8\% for $\mathtt{CCPHASE_{011}}(\pi/2)$. Note that the QPT data was generated using a calibration procedure for $\mathtt{CCPHASE_{011}}$ that included an additional step for calibrating the conditional phase applied by the intermediate $\mathtt{CPHASE}$ on $|q_1 q_2\rangle$; however, subsequent benchmarking data in this report was gathered without explicitly calibrating this phase, as the inter-pulse phase for the source $\mathtt{CPHASE}$ gate is calibrated during routine standard recalibration procedures for the Aspen-9 system. See Appendix~\ref{section:quilt_calibration} for details.

\subsection{Benchmarking of Arbitrary Doubly-Controlled Phase Gates}
\label{section:results}

Rigorous benchmarking of three-qubit gates is considerably more computationally expensive than two-qubit gate benchmarking in terms of both classical and quantum resources. Furthermore, because the $\textrm{CCZ}$ is located in the third level of the Clifford hierarchy~\cite{Cui_2017,Bravyi_2019}, it cannot be benchmarked using standard randomized-benchmarking protocols \cite{Knill2008, Magesan2012}. This presents a challenge for quantifying performance in these multi-qubit spaces.

To evaluate the performance of the calibrated gate for a variety of input angles, we benchmarked the $\mathtt{CCPHASE_{011}}$ gate on the Aspen-9 sublattice (10, 11, 12) using Cycle Benchmarking (CB) \cite{Erhard2019, Beale2020True-Q} . Cycle Benchmarking is a scalable randomized benchmarking protocol that permits benchmarking of non-Clifford gates, at the cost of a possibly expensive correction (in terms of three-qubit gate count) at the end of each sequence. When this correction is used for non-Clifford gates such as $\mathtt{CCPHASE_{011}}$, CB may produce a noisy estimate of the fidelity, as the final inversion may require many three-qubit gates. Nevertheless, using CB we were able to benchmark significantly more calibrations and output conditions (target phases) than would have been possible using QPT.

A point test of the textbook $\mathtt{CCPHASE}$ gate was conducted as $\mathtt{CCPHASE}(\pi)=\mathtt{CCZ}$ using CB, producing an estimated process fidelity of $\mathcal{F}_{\mathtt{CCZ}}=87.1 \pm 0.8\%$. Further, a point test of a $\mathtt{CCNOT}$ gate constructed from the native three-qubit gate was conducted using CB, producing an estimated process fidelity of $\mathcal{F}_\mathtt{CCNOT}=85.1 \pm 0.8\%$. These values are consistent with the observed performance of $\mathtt{CCPHASE_{011}}$ in the presence of additional 1Q gates.

\begin{figure}
     \centering
     (a)\includegraphics[scale=0.6, trim={ 0.0cm  0 0cm 0},clip]{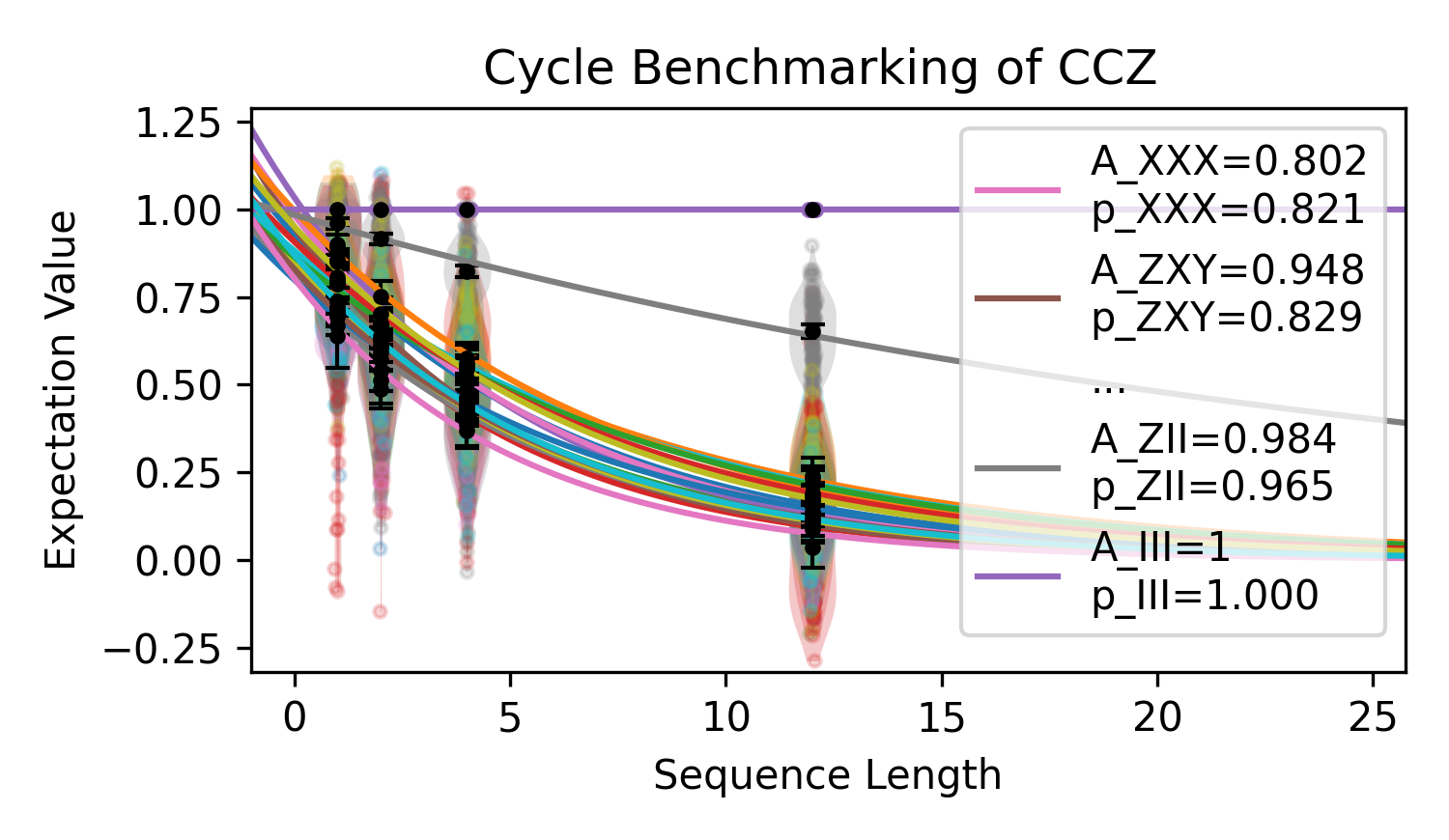}
     (b)\includegraphics[scale=0.6, trim={ 0.0cm 0  0cm 0},clip]{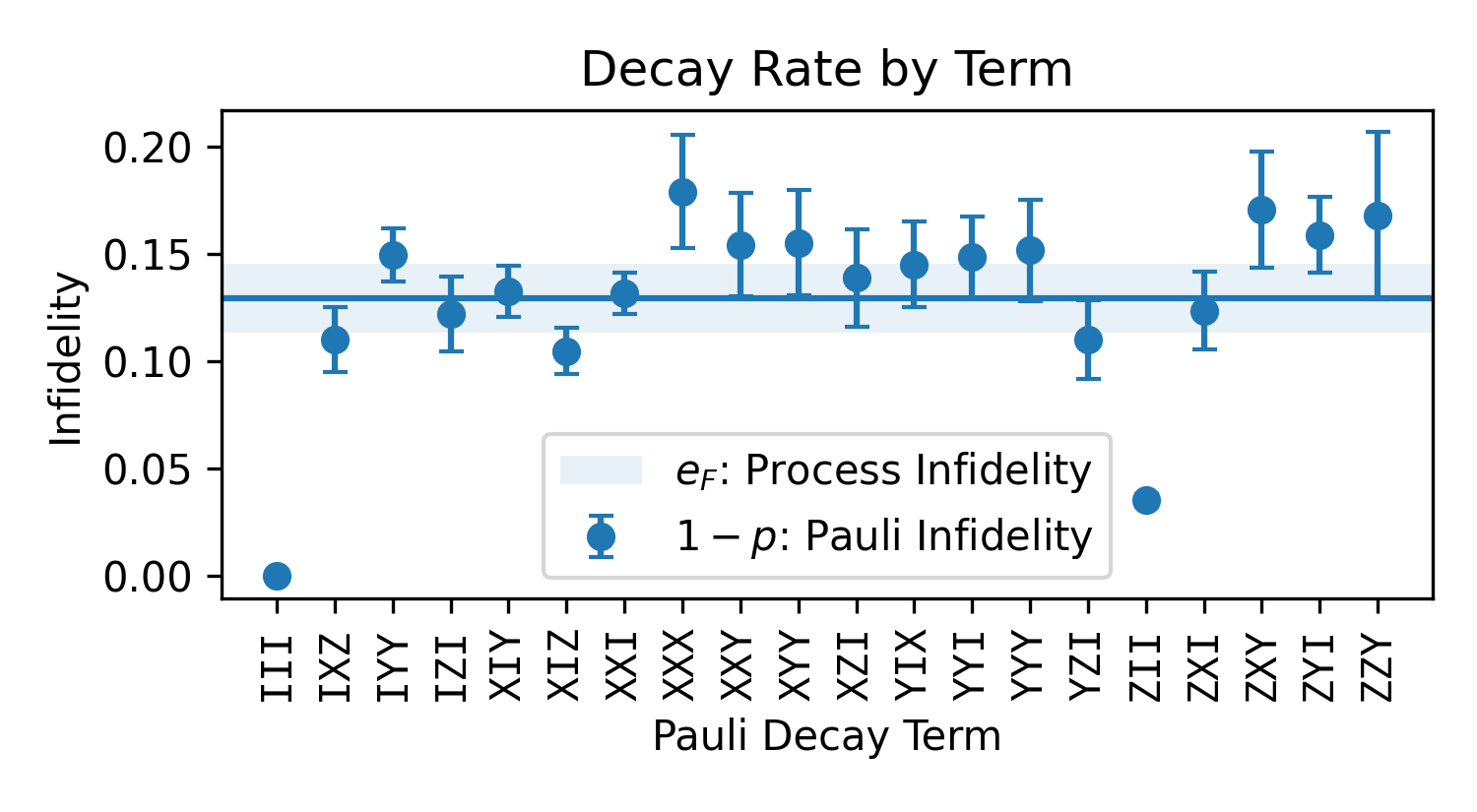}
     (c)\includegraphics[scale=0.55, trim={ 3cm 0  3cm 0},clip]{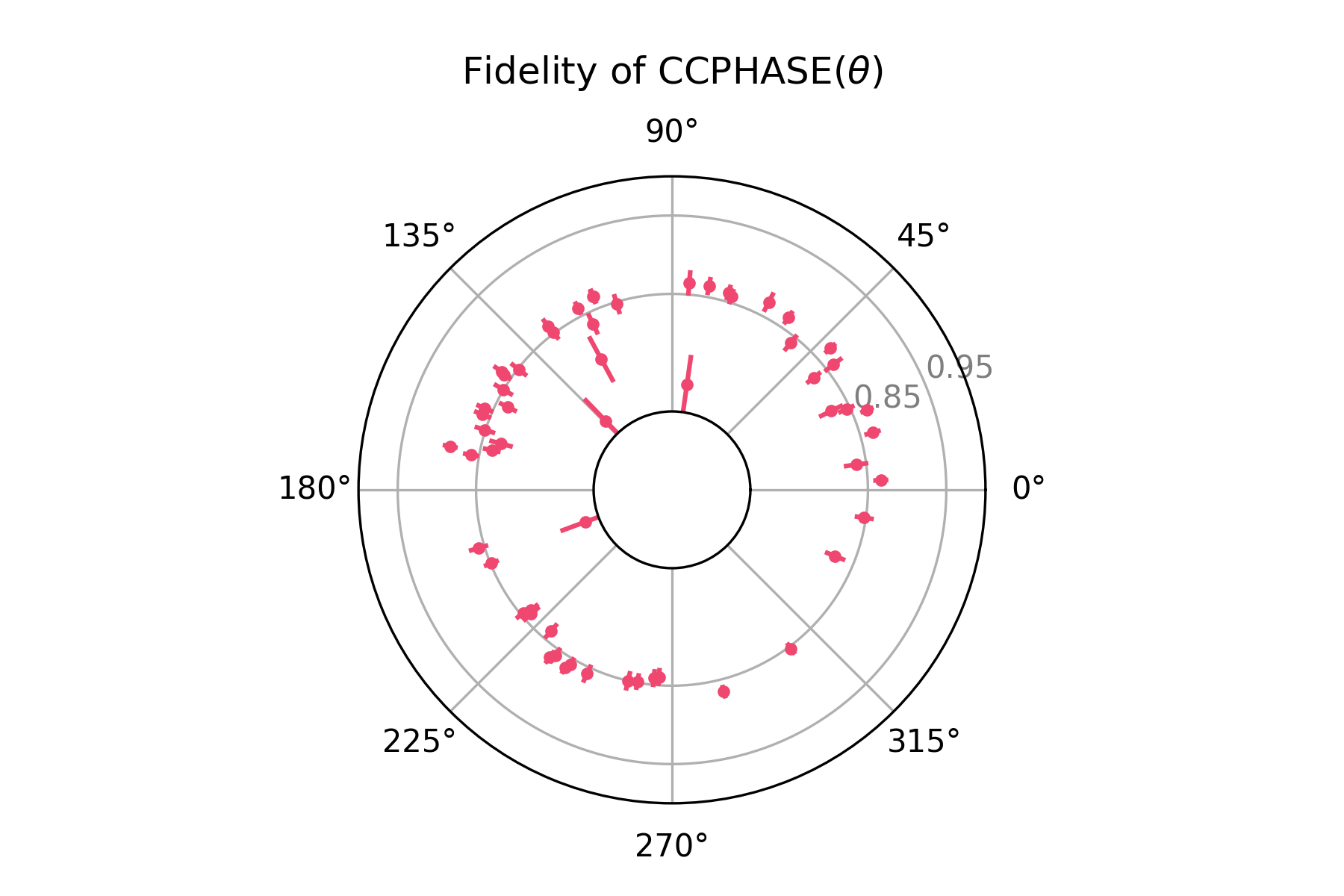}
     \caption{Results of Cycle Benchmarking. (a) CB decay fits and (b) associated decay rates for individual Pauli decay terms of interest. (c) Benchmarking of $\mathtt{CCPHASE_{011}}$ with respect to input conditional phase angle on Aspen-9 sublattice (10, 11, 12); median 85.2\%. Results are consistent for arbitrary angles.}
     \label{fig:cb_angles}
 \end{figure}

When benchmarking the performance of arbitrary angles, we instead benchmark the composite gate $\mathtt{CCPHASE_{011}}(\theta)$  $\mathtt{CCPHASE_{011}}(2\pi-\theta)$, equivalent to the identity. The fidelity for each individual gate is then bounded by the square root of the fidelity of the composite gate. This method increases the error in the estimate of the per-gate process fidelity, but eliminates the penalty incurred in the final non-Clifford correction step. We additionally benchmark $\mathtt{CCPHASE_{011}}(\pi)=\mathtt{CCZ}_{011}$ to verify that the conditional phase matches the requested value. Figure~\ref{fig:cb_angles} shows the results of Cycle Benchmarking of the qutrit-based $\mathtt{CCPHASE_{011}}$ gate. We achieved a median fidelity of 85.2\% over 58 randomized conditional phases, with no observed dependence of fidelity on angle. A small number of outlying angles were observed in the results, which we attribute to intermittent degradation in readout performance for one qubit of the triplet.

To provide value over the two-qubit decomposition, the three-qubit gate must have a fidelity that exceeds the fidelity of the entire two-qubit decomposition, including single-qubit gates. We can directly compare the performance of the native three-qubit gate to the full two-qubit decomposition by benchmarking the entire equivalent circuit via Cycle Benchmarking. We discuss one such decomposition in Appendix~\ref{section:reference_decomposition}, which we take as the two-qubit decomposition our native qutrit implementation should outperform. Given the topological constraints of the Aspen-9 platform, this decomposition results in 3 $\mathtt{CPHASE}$ + 6 $\mathtt{CNOT}$ gates. We benchmarked this decomposition on the same set of three qubits as before using Cycle Benchmarking, producing an estimated process fidelity of $76 \pm 5\%$. We provide this figure as an upper bound, as CB of the full decomposition produced a number of Pauli decay terms with a vanishing expectation value at a sequence depth of just 2 cycles, suggesting that the true error may be too high to benchmark accurately. As discussed in \Cref{section:reference_decomposition}, under reasonable assumptions of the performance of the source gates, we would anticipate a maximum fidelity of 75.0\% for the full decomposition based on single- and two-qubit gates in the absence of correlated gate errors -- in rough agreement with the CB value. By this metric, our qutrit-based implementation significantly outperforms the full native decomposition of $\mathtt{CCPHASE}$.

We note, however, that the benchmark results for $\mathtt{CCPHASE}$ are lower than would be expected by simply multiplying out the fidelities of the donor $\mathtt{CPHASE}$ gates providing the source $\mathtt{iSWAP_{02/20}}$ pulses -- roughly $97.7\%\times97.3\%= 95.1\%$ for edges (10,11) and (11,12), respectively. The use of the second excited state of the transmon with its relatively poor $|1\rangle \leftrightarrow |2\rangle$ coherence is likely the cause of part of the discrepancy. This has been verified in part by measuring the $|1\rangle\leftrightarrow|2\rangle$ dephasing time, which was found to be significantly lower than the $|0\rangle\leftrightarrow|1\rangle$ dephasing time (especially for fixed-frequency qubits, see Table~\ref{table:qubit_properties}). 

Simulations of the impact of dephasing and relaxation of the first two excited states on the $\mathtt{CCPHASE}$ gate fidelity yielded an estimated coherence limit of 92.8\%, assuming an otherwise ideal Hamiltonian and a total effective gate time of 402 ns (including pulse padding and risetime). When including the contribution of modulation-induced dephasing (typically reducing the bare dephasing time by approximately 50\% for qubits 10 and 12) we estimate a coherence-limited fidelity bound of approximately 87\%. This indicates that the error in our gate is primarily the result of decoherence, and that reducing the impact of charge noise on dephasing by increasing the ratio $E_J/E_C$ of each transmon qubit or mitigating the effects of AC flux noise would likely improve the performance of these gates.

\subsection{Applications to Constraint Satisfaction}
\label{section:qaoa}

We now turn our attention to demonstrating the important role of $\mathtt{CCPHASE}$ in solving MAX-3-SAT with QAOA. First, we provide a description of the MAX-3-SAT problem. Then, we review how to apply QAOA to the MAX-3-SAT task, explaining how the construction requires a unitary that is equivalent to the $\mathtt{CCPHASE}$ gate. Finally, we report a small-scale experiment implementing a single MAX-3-SAT clause on three qubits, deploying the $\mathtt{CCPHASE}$ to achieve quantitative agreement with theory.

\begin{figure*}
    \centering
    \includegraphics[scale=.8]{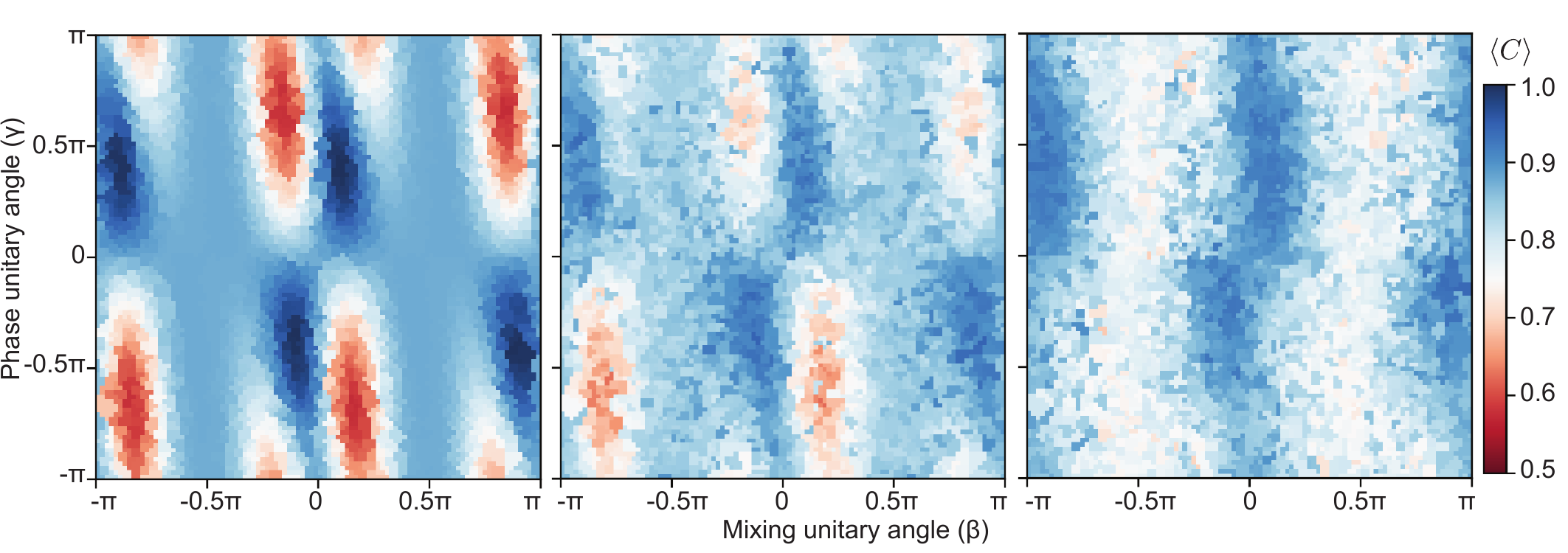}
    \caption{Cost function landscape of MAX-3-SAT via QAOA. Quantum circuit construction for the clause $c=\left(x_0 \vee x_1 \vee x_2 \right)$ is three steps: transversal Hadamard gates, phase separation for $\mathtt{CCPHASE}_{111}(\gamma)$, and mixing via transversal Pauli-$X$ rotation by $\beta$. Measurements are done in the $Z$ basis to compute $\langle C \rangle$ as discussed in the main text. (left) Ideal output of the QAOA ansatz for a noiseless quantum circuit simulator. (middle) Data from the quantum processor using the native $\mathtt{CCPHASE}_{111}(\gamma)$. (right) Data from the quantum processor compiling $\mathtt{CCPHASE}_{111}(\gamma)$ to a native two-qubit gate set.
}
    \label{fig:qaoa_fig}
\end{figure*}

The MAX-3-SAT problem statement is defined as follows. Let $c$ be a set of $m$ disjunctive clauses: a collection of Boolean logic formulas based on the OR operator $\vee$. Further, let each clause contain three of a possible $n$ Boolean problem variables ($\mathbf{x}$) or their logical negation (literals). We denote $\mathbf{x}_{k}$ as the state of the variables contained in clause $c_k$. We say $c_k$ is satisfied if the logical formula evaluates to 1: $c_{k}(\mathbf{x}_{k})=1$. Otherwise, the clause evaluates to zero. Thus,  the total number of satisfied clauses is then: $c(\mathbf{x})=\sum_{k=1}^{m}c_{k}(\mathbf{x}_{k})$. The objective of MAX-3-SAT is to maximize $c(\mathbf{x})$.

This can be approached using QAOA. The algorithm requires three distinct components. First, an initial state must be prepared with support over the target solutions. For the standard QAOA design, that is accomplished with a single round of transversal Hadamard gates applied across an $n$-qubit register, generating an equal weight of all $n$-bit strings. That is an appropriate choice for MAX-3-SAT, too, since the resulting distribution includes the optimal bit string. 

Second, a phase separation operator is required, given by exponentiation of the diagonal cost function Hamiltonian, which for MAX-3-SAT is given by \cite{Farhi2014}, Eq.~(2): 
\begin{equation}
    U_C(\gamma)=\prod_{k=1}^{m}e^{-i\gamma C_{k}}
\end{equation}
with $C_k$ the quantum operator that encodes original scalar cost function $c_k$ into Pauli spin operators. To motivate that encoding, we first convert the problem from the Boolean to spin algebra via the trivial isomorphism: $\mathbf{s}=1-2\mathbf{x}$, consistent with the eigenvalues of the Pauli-$Z$ operator. We can then write the classical scalar cost function for a single clause in the general form:
\begin{equation}
\label{eq:scalar}
c_{k} = 1 - \tfrac{1}{8} \left( 1 \pm s_{k,0} \right) \left( 1 \pm s_{k,1} \right) \left( 1 \pm s_{k,2} \right)
\end{equation}
where $s_{k,\ell}\in\{-1,1\}$ is the $\ell$-th variable of the $k$-th constraint clause, and the sign convention is determined with $\pm$ taken as $+$ in the case that the associated logical variable is not negated in $c_k$ (positive literal) or $-$ in the case of negation (negative literal). There are eight possible clauses over three unique literals ($2^3$). As an example, consider the all-positive literal clause:
\begin{equation}
\left(x_0 \vee x_1 \vee x_2 \right) \mapsto 1 - \tfrac{1}{8} \left( 1 + s_0 \right) \left( 1 + s_1 \right) \left( 1 + s_2 \right)
\end{equation} 
which has the corresponding truth table:
\begin{equation}
\begin{matrix}
x_0 & x_1 & x_2 & s_0 & s_1 & s_2 & x_0 \vee x_1 \vee x_2 \\ 
0 & 0 & 0 & +1 & +1 & +1 & 0 \\ 
0 & 0 & 1 & +1 & +1 & -1 & 1 \\ 
0 & 1 & 0 & +1 & -1 & +1 & 1 \\
0 & 1 & 1 & +1 & -1 & -1 & 1 \\ 
1 & 0 & 0 & -1 & +1 & +1 & 1 \\
1 & 0 & 1 & -1 & +1 & -1 & 1 \\ 
1 & 1 & 0 & -1 & -1 & +1 & 1 \\
1 & 1 & 1 & -1 & -1 & -1 & 1 \\ 
\end{matrix}
\end{equation}
Note that each of the eight unique clauses correspond to a distinct row for a zero-valued, unsatisfied constraint evaluation. To quantize the general form of $c_k$ from Eq.~\ref{eq:scalar}, we simply substitute $Z_{k,\ell}$ for $s_{k,\ell}$, as
\begin{equation}
C_{k} = 1 - \tfrac{1}{8} \left( 1 \pm Z_{k,0} \right) \left( 1 \pm Z_{k,1} \right) \left( 1 \pm Z_{k,2} \right)
\end{equation}
using the same sign convention for positive and negative literals as before. Analogously to the previous truth table, we have $C_{k}|\mathbf{x}_k\rangle=|\mathbf{x}_k\rangle$ for all except one state, which does not satisfy the constraint and gives $C_{k}|\mathbf{x}_k\rangle=0$. Therefore, the phase separation operator is a diagonal matrix whose entries are all $\exp{(-i\gamma)}$ except for one element, corresponding to the unsatisfied state, whose value is 1. For instance, our previous example clause has
\begin{equation}
U_C(\gamma) = e^{-i\gamma}\text{diag}\left(e^{i\gamma},1,1,1,1,1,1,1\right).
\end{equation}
Importantly, this unitary is equivalent to $\mathtt{CCPHASE}_{000}(\gamma)$. Indeed, all of the other unique clauses are accessible by a single $\mathtt{CCPHASE}$ gate with bit flips (Pauli-$X$ gates) to align the literals before and afterwards. Thus, to accumulate a total phase separation operator over $m$ clauses, at most $m$ $\mathtt{CCPHASE}$ gates are required.

The last component required for our QAOA construction is a mixing unitary, which interferes the register states after the phase separation step. Once again, because all $n$-bit strings represent feasible solutions to any problem instance, the standard QAOA mixing operator $U_B(\beta)=\prod_{j=1}^{n}e^{-i\beta \sigma_x^{j}}$ will suffice. This can be implemented with transveral single-qubit Pauli-$X$ rotations.

After performing initial state preparation, phase separation, and mixing steps, the qubit register can be measured to extract one $n$-bit string sample. With classical post-processing, the scalar value $c(\mathbf{x})$ can be determined for each sample over multiple repetitions of the same configuration. Further, a variety of parameter setting and learning strategies can find values of ($\beta$,$\gamma$) that maximize $c(\mathbf{x})$. In principle, the quality of solutions obtained by QAOA is also improved by repeating alternating layers of phase separation and mixing unitaries with independent variational parameters before measurement.

With all of the necessary building blocks, we now consider the experimental implementation of QAOA for MAX-3-SAT (see Fig.~\ref{fig:qaoa_fig}). We take $C=x_0 \vee x_1 \vee x_2$, the example clause from before as our problem instance, meaning that the phase separation unitary is $\mathtt{CCPHASE}_{000}(\gamma)$. We achieve this operation via the native $\mathtt{CCPHASE}_{011}$ gate by compiling bit flips to the second and third qubits before and after the application of the phase gate. We also considered a version of the ansatz compiled to a two-qubit gate set, which included native $\mathtt{CCPHASE}$ gates. Both tests comprise 2500 samples per configuration of ($\beta$,$\gamma$) and through a random search of 2500 configurations of $(\beta,\gamma)\in[0,\pi]^{\otimes 2}$, with the same variational parameters tested in a noiseless simulator and both circuit constructions. The expectation value of $\langle C \rangle$ is computed for each configuration.

Taking the whole cost function landscape into consideration, we compute the linear correlation comparing the ideal algorithm to observed QAOA output on hardware, for both native and compiled phase separation operators. We find a nearly four-fold improvement in the Pearson correlation coefficient for the native case ($\rho=0.838$ versus $\rho=0.225$), demonstrating the more faithful construction of the QAOA MAX-3-SAT algorithm with the three-qubit gate. Moreover, we find quantitative improvement to the penalty term $\mu\equiv\textrm{min}(\langle C(\beta,\gamma) \rangle$) found for ($\beta$,$\gamma$)=(2.643,-2.055): $\mu=0.738$ for $\mathtt{CCPHASE}$, versus $\mu=0.894$ for two-qubit gates; compared to $\mu=0.576$ for noiseless simulations and $\mu=7/8$ for sampling random 3-bit strings. Finally, it is notable that the gradient is preserved for the $\mathtt{CCPHASE}$ case, which is essential for hardware-trained parameter setting strategies inherit to problem instances beyond classical HPC simulators.

\section{Conclusion}
We have demonstrated a three-qubit, doubly-controlled $\mathtt{CCPHASE}$ gate with an arbitrary conditional phase on our Aspen-9 platform that outperforms decomposition using local, two-qubit gates. We applied this gate to enable an algorithm with a path to real-world quantum advantage, MAX-3-SAT, and demonstrated a qualitative and quantitative improvement in the generated QAOA landscape when using the custom gates over the full two-qubit decomposition. 

This experiment was conducted via an open source language for pulse-level control, Quil-T~\cite{Quil-TDocumentation}, which demonstrates the flexibility of emerging platforms for performing circuit-, gate-, and pulse-level optimizations. Access to the low-level pulse programs enacting each gate unlocks the ability to augment existing gates: for example, building echo sequences into existing gates can provide insensitivity to certain types of noise \cite{Sundaresan2020}. Existing gate structures can also be repurposed -- in this work, for example, we demonstrated how existing $\mathtt{CPHASE}$ gates can be decomposed into reusable two-qutrit primitives. The broad availability of these two-qutrit operations can facilitate research well beyond the work presented here, allowing for the development of sophisticated approaches to circuit depth reduction \cite{Gokhale2019}, exploration of high-dimensional entanglement \cite{CerveraLierta2021}, and the study of qubit and qutrit quantum dynamics \cite{Blok2021}.

\acknowledgements
We thank Keysight Technologies for helpful discussions and guidance in using their Cycle Benchmarking toolkit~\cite{Beale2020True-Q}; Joseph Valery and Stefano Poletto for valuable feedback on this manuscript. We thank Bram Evert for software infrastructure support. We thank Davide Venturelli and Eleanor Rieffel for insightful conversations on QAOA.  A.D.H. developed the calibration and benchmarking procedures. N.D. provided the theory support for gate physics. M.J.R. and M.J.H. identified and developed the QAOA application. A.D.H. and M.J.R. wrote the manuscript with inputs from all authors. The authors declare no competing interests. The experimental results presented here are based
upon work supported by the Defense Advanced Research Projects Agency (DARPA) under agreement
No. HR00112090058.

\appendix

\section{Reference decomposition}
\label{section:reference_decomposition}
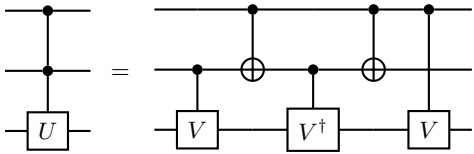
\begin{figure}[h]
\centering
\begin{quantikz}[row sep={0.8cm,between origins}, column sep=0.3cm]
    \qw & \ctrl{1} & \qw \\
    \qw & \ctrl{1} & \qw \\
    \qw & \gate{U} & \qw 
    \end{quantikz} $\,$
= \begin{quantikz}[row sep={0.8cm,between origins}, column sep=0.3cm]
    \qw & \qw      & \ctrl{1} & \qw               & \ctrl{1} & \ctrl{2} & \qw \\
    \qw & \ctrl{1} & \targ{}  & \ctrl{1}          & \targ{}  & \qw      & \qw \\
    \qw & \gate{V} & \qw      & \gate{V^\dagger}  & \qw      & \gate{V} & \qw 
\end{quantikz}
\caption{Decomposition of controlled-controlled-$U$ gate, as in Ref.~\protect\cite{Barenco1995}.}
\label{fig:ccphase_decomposition_1}
\end{figure}

The usefulness of our native representation of $\mathtt{CCPHASE}$ rests on its ability to outperform decompositions involving only standard two-qubit gates, such as XY and CPHASE. Here we develop a reference decomposition for $\mathtt{CCPHASE}$ against which we can benchmark our native qutrit implementation. We only consider decompositions which provide arbitrary phase control with a fixed circuit (in the case where $\mathtt{CCPHASE}(\theta = 0) = \mathbf{I}$, the reference decomposition is trivial but not general). Ref.~\cite{Barenco1995} provides the basis of our $\mathtt{CCPHASE}$ reference decomposition, illustrated in \Cref{fig:ccphase_decomposition_1}. It defines the decomposition of a 3Q controlled-controlled-unitary $U$ in terms of 2Q controlled-unitary $V$ and 2Q controlled-unitary $V^\dagger$, with $V^2=U$, and $\mathtt{CNOT}$ gates.

In our case, we set
\begin{equation*}
    U = \textrm{PHASE}(\theta) = \begin{pmatrix}
      1 & 0 \\
      0 & e^{i \theta}
    \end{pmatrix}\textrm{,}
\end{equation*}
so that controlled-controlled-$U$ is our desired three-qubit (3Q) gate:

\begin{equation*}
    \textrm{CCPHASE}(\theta) = \begin{pmatrix}
      1 & 0 & 0 & 0 & 0 & 0 & 0 & 0 \\
      0 & 1 & 0 & 0 & 0 & 0 & 0 & 0 \\
      0 & 0 & 1 & 0 & 0 & 0 & 0 & 0 \\
      0 & 0 & 0 & 1 & 0 & 0 & 0 & 0 \\
      0 & 0 & 0 & 0 & 1 & 0 & 0 & 0 \\
      0 & 0 & 0 & 0 & 0 & 1 & 0 & 0 \\
      0 & 0 & 0 & 0 & 0 & 0 & 1 & 0 \\
      0 & 0 & 0 & 0 & 0 & 0 & 0 & e^{i \theta}
    \end{pmatrix}\textrm{,}
\end{equation*}
while controlled-$V$ and controlled-$V^\dagger$ are members of the $\mathtt{CPHASE}$ gate family, with arguments $\tfrac{\theta}{2}$ and $-\tfrac{\theta}{2}$, respectively.

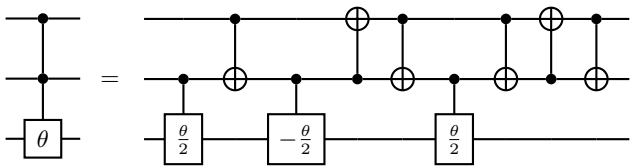
\begin{figure}[h]
\centering
\begin{quantikz}[row sep={0.8cm,between origins}, column sep=0.275cm]
    \qw & \ctrl{1}      & \qw \\
    \qw & \ctrl{1}      & \qw \\
    \qw & \gate{\theta} & \qw 
    \end{quantikz} $\,$
= \begin{quantikz}[row sep={0.8cm,between origins}, column sep=0.275cm]
    \qw & \qw                      & \ctrl{1} & \qw                        & \targ{}   & \ctrl{1} & \qw                      & \ctrl{1} & \targ{}   & \ctrl{1} & \qw \\
    \qw & \ctrl{1}                 & \targ{}  & \ctrl{1}                   & \ctrl{-1} & \targ{}  & \ctrl{1}                 & \targ{}  & \ctrl{-1} & \targ{}  & \qw \\
    \qw & \gate{\tfrac{\theta}{2}} & \qw      &  \gate{-\tfrac{\theta}{2}} & \qw       & \qw      & \gate{\tfrac{\theta}{2}} & \qw      & \qw       & \qw      & \qw
\end{quantikz}
\caption{Decomposition of $\mathtt{CCPHASE}$ gate to local $\mathtt{CPHASE}$ and $\mathtt{CNOT}$ gates.}
\label{fig:ccphase_decomposition_2}
\end{figure}

Our physical implementation of $\mathtt{CCPHASE}$ requires only linear chain connectivity between the three qubits. The final controlled-$V$ operation in the reference decomposition of \Cref{fig:ccphase_decomposition_1} is therefore non-local, and not physically realizable by $\mathtt{CPHASE}$ gates implemented on our superconducting quantum processor \cite{Abrams2020}. We inject $\mathtt{SWAP}$ operations to address this non-locality, then substitute and simplify to $\mathtt{CNOT}$ gates \cite{Garcia-Escartin2011EquivalentCircuits} to realize the $\mathtt{CNOT}$-based decomposition of \Cref{fig:ccphase_decomposition_2}.

\begin{figure}[h]
\centering
\begin{quantikz}[row sep={0.8cm,between origins}, column sep=0.3cm]
    \qw & \ctrl{1} & \qw \\
    \qw & \targ{}  & \qw 
    \end{quantikz} $\,$
= \begin{quantikz}[row sep={0.8cm,between origins}, column sep=0.3cm]
    \qw & \qw      & \ctrl{1} & \qw \\
    \qw & \gate{H} & \gate{Z} & \gate{H}
\end{quantikz}

\begin{quantikz}[row sep={0.8cm,between origins}, column sep=0.3cm]
    \qw & \gate{H} & \qw
    \end{quantikz} $\,$
= \begin{quantikz}[row sep={0.8cm,between origins}, column sep=0.3cm]
    \qw & \gate{\textrm{R}_z(\tfrac{\pi}{2})} & \gate{\textrm{R}_x(\tfrac{\pi}{2})} & \gate{\textrm{R}_z(\tfrac{\pi}{2})} & \qw
\end{quantikz}
\caption{Decomposition of CNOT gate to R$_x$, R$_z$, and CZ gates.}
\label{fig:ccphase_decomposition_3}
\end{figure}
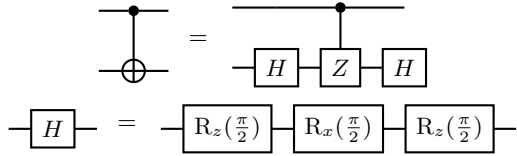

Our final step is to decompose $\mathtt{CNOT}$ gates into $\mathtt{RX}$, $\mathtt{RZ}$, and $\mathtt{CZ}$ gates. This follows \cite{Garcia-Escartin2011EquivalentCircuits} and is illustrated in \Cref{fig:ccphase_decomposition_3}. The result is a $\mathtt{CCPHASE}$ reference decomposition that is realizable on our superconducting quantum processor using the existing gate set of $\mathtt{CPHASE}$, $\mathtt{CZ}$, $\mathtt{RX}$, and $\mathtt{RZ}$. There are 9 two-qubit (2Q) gates and 12 non-trivial\footnote{$\mathtt{RZ}$ is implemented trivially on our superconducting quantum processors, consisting only of a frame shift in our control systems. It therefore has negligible impact on circuit time, and errors in the applied phase can be considered as accumulating into each real $\mathtt{RX}$ gate \cite{McKay2017}.} one-qubit (1Q) gates in this decomposition. Anticipating a 1Q gate fidelity of 99.5\% and a 2Q gate fidelity of 97\%, we estimate the fidelity of the decomposed 1Q+2Q $\mathtt{CCPHASE}$ operator to be $0.975^{9} \times 0.995^{12} \approx 75\%$. This provides a benchmark fidelity estimate against which our physical implementation of $\mathtt{CCPHASE}$ can be evaluated.

\section{Calibration of CCPHASE using Quil-T}
\label{section:quilt_calibration}

Quil-T is an extension to the Quil instruction language for quantum computers that allows pulse-level programming of our superconducting qubit-based quantum processors. All native gates available on our deployed processors provide a Quil-T description that fully dictates the underlying physical behavior of the gate. For more information, see Ref.~\cite{Quil-TDocumentation}. All results in this report are producible using Quil-T and pyQuil over our Quantum Cloud Services (QCS) platform by manipulating and rearranging the existing pulse parameters for deployed $\mathtt{CCPHASE}$ gates on Aspen-9.

The qutrit-based 3Q gate $\mathtt{CCPHASE_{011}}$ is obtained by deconstructing the deployed $\mathtt{CPHASE}$ gate on the $|q_0 q_1\rangle$ edge into the two constituent parts, and calibrating the necessary constituent phases. The general pulse program for $\mathtt{CCPHASE_{011}}$ can be expressed in Quil-T with $\langle\rangle$ placeholders for qubit identifiers and phases to be calibrated:

\begin{verbatim}
DEFCAL iSWAP[02|20]_P1 <q0> <q1>
  FENCE
  NONBLOCKING PULSE <q0> <q1> "cphase" q0/q1_iSWAP02
  FENCE

DEFCAL iSWAP[02|20]_P2 <q0> <q1>
  FENCE
  SHIFT-PHASE <q0> <q1> "cphase" +-0.5*<B>
  NONBLOCKING PULSE <q0> <q1> "cphase" q0/q1_iSWAP02
  SHIFT-PHASE <q0> <q1> "cphase" -+0.5*<B>
  FENCE

DEFCAL CCPHASE011(<A>) <q0> <q1> <q2>:
  iSWAP[02|20]_P1 <q0> <q1>
  CPHASE(<A>) <q1> <q2>
  iSWAP[02|20]_P2 <q0> <q1>
  RZ(<C>) <q0>
  RZ(<D>) <q1>
  RZ(<E>) <q2>
\end{verbatim}

\begin{figure}
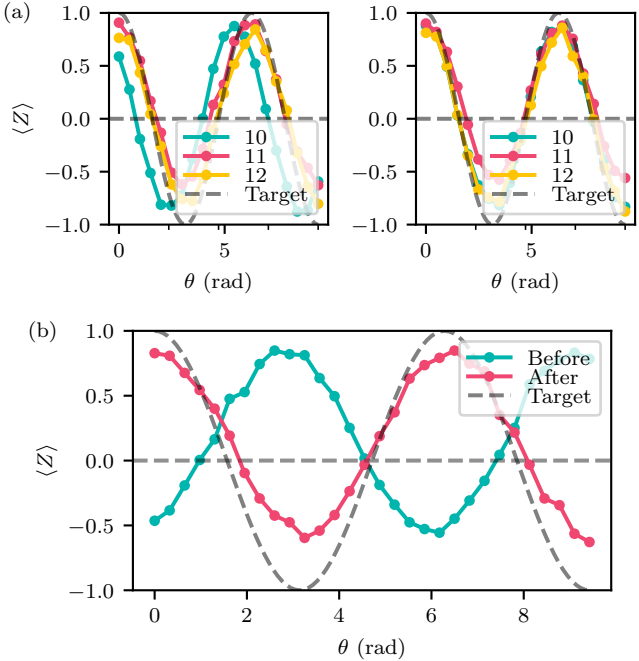

     \centering
    \begin{adjustbox}{trim=0.5cm 0.25cm 0cm -0.5cm}
    \input{rzs_before_after.pgf}
    \end{adjustbox}
    \begin{adjustbox}{trim=0.5cm 0.5cm 0cm 0cm}
        \input{identity_calibration.pgf}
    \end{adjustbox}
     \caption{Tune up of individual phases for the $\mathtt{CCPHASE}$ gate. (a) Single-qubit phases \{C, D, E\} on Aspen-9 qubits (10,11,12) before (left) and after (right) calibration. (b) Conditional phase $B$ of the (10, 11) subspace as observed on 11 before and after calibration.}
     \label{fig:phase_calibration}
 \end{figure}

There are five phases that require calibration: $A$, the conditional phase, set so that the three-qubit conditional phase matches the requested value; $B$ the phase of the final $\mathtt{iSWAP_{02/20}}$ pulse, set so the final unitary is identity when the conditional phase is zero ; \{$C , D, E$\} single-qubit phase rotations that cancel any phases accumulated during the previous flux operations. Phase $A$ is calibrated during the $\mathtt{CPHASE}$ calibration procedure deployed on Aspen-9, so only phases \{$B,C,D,E$\} need explicit treatment for $\mathtt{CCPHASE_{011}}$.

\{$C, D, E$\} The single qubit phases are almost independent of the phase $B$ on the flux frame. These are calibrated using a Quil-T Ramsey interferometry measurement, with each \texttt{q0}, \texttt{q1}, \texttt{q2} assessed independently:
\begin{verbatim}
RX(+pi/2) <qx>
CCPHASE(0) <q0> <q1> <q2>
RZ(<Ramsey Phase>)
RX(-pi/2) <qx>
MEASURE <qx>
\end{verbatim}

We fit the phase offset of the oscillations in Z to cancel the phase accumulated by each qubit during the $\mathtt{CCPHASE_{011}}$ operation (Figure~\ref{fig:phase_calibration}(a)).

\{B\} The unitary for the subspace formed by $|q_0 q_1\rangle$ should be identity when $|q_2\rangle=|0\rangle$. We tune phase \{B\} so that all terms in the three-qubit unitary are equivalent to the identity apart from the conditional phase. We repeat a Ramsey-interferometric measurement, but examine the phase on \{q1\}:
\begin{verbatim}
RX(pi) <q0>
RX(+pi/2) <q1>
CCPHASE(0) <q0> <q1> <q2>
RZ(<Ramsey Phase>)
RX(-pi/2) <q1>
MEASURE <q1>
\end{verbatim}

This measures the conditional phase of the $\mathtt{CPHASE}$ gate on $|q_0 q_1\rangle$ formed by the first and last $\mathtt{iSWAP}$ pulses in \Cref{fig:overview_fig}, but potentially affected by the intervening pulses on $|q_1 q_2\rangle$. The phase on the final $\mathtt{iSWAP_{02/20}}$ flux pulse $B$ is set so that the conditional phase in this subspace is zero (Figure~\ref{fig:phase_calibration}(b)).

The single-qubit phases \{C, D, E\} may not be completely independent of the conditional phase \{B\} due to small coherent errors introduced by the finite risetime of the constituent pulses. This effect can be mitigated by performing two passes of the calibration procedure at a target angle of interest.

\section{Allowed combinations of iSWAP interactions for CCPHASE}
\label{section:allowed_combinations}
\newcommand{\iSWAP}[1]{\gate[2, style={inner ysep=-20pt, inner xsep=4pt}]{\rotatebox{90}{$\mathtt{iSWAP}_{02/20}(#1)$}}}

\begin{table*}[ht]
\begin{quantikz}[row sep={1.5cm,between origins}]
q_0 & \iSWAP{0} \slice{$t_1$} & \qw \slice{$t_2$}& \qw \slice{$t_3$}& \iSWAP{\pi} \slice{$t_4$}& \qw\\ 
q_1 &  & \iSWAP{0} & \iSWAP{\theta} & &\qw \\
q_2 & \qw & & \qw &\qw & \qw
\end{quantikz}
\begin{minipage}{.5\linewidth}
\begin{tabular}{c | c c c c c}
Pulse Type & Input  & $t_1$ & $t_2$ & $t_3$ & $t_4$ \\
\hline
  iSWAP02,     & $|011\rangle$ & $|011\rangle$ & $-i|002\rangle$ & $e^{-i\theta}|011\rangle$    & $e^{-i\theta}|011\rangle$\\
  iSWAP02 & $|110\rangle$ & $-i|020\rangle$ & $-i|020\rangle$ & $-i|020\rangle$           & $|110\rangle$ \\
    & $|111\rangle$ & $-i|021\rangle$ & $-i|021\rangle$ & $-i|021\rangle$           & $|111\rangle$ \\ \hline
 iSWAP02,       & $|011\rangle$ & $|011\rangle$ & $-i|020\rangle$ & $e^{-i\theta}|011\rangle$    & $e^{-i\theta}|011\rangle$ \\
 iSWAP20& $|110\rangle$ & $-i|020\rangle$ & $-|011\rangle$ & $-ie^{-i\theta}|020\rangle$ & {\color{red}{$e^{-i\theta}|110\rangle$}} \\
      & $|111\rangle$ & $-i|021\rangle$ & $-i|021\rangle$ & $-i|021\rangle$           & $|111\rangle$ \\ \hline
  iSWAP20,     & $|011\rangle$ & $|011\rangle$ & $-i|002\rangle$ & $e^{-i\theta}|011\rangle$    & $e^{-i\theta}|011\rangle$ \\
 iSWAP02& $|110\rangle$ & $-i|200\rangle$ & $-i|200\rangle$ & $-i|200\rangle$           & $|110\rangle$ \\
      & $|111\rangle$ & $-i|201\rangle$ & $-i|201\rangle$ & $-i|201\rangle$           & $|111\rangle$ \\ \hline
 iSWAP20,       & $|011\rangle$ & $|011\rangle$ & $-i|020\rangle$ & $e^{-i\theta}|011\rangle$    & $e^{-i\theta}|011\rangle$ \\
 iSWAP20& $|110\rangle$ & $-i|200\rangle$ & $-i|200\rangle$ & $-i|200\rangle$           & $|110\rangle$ \\
      & $|111\rangle$ & $-i|201\rangle$ & $-i|201\rangle$ & $-i|201\rangle$           & $|111\rangle$ \\
\end{tabular}
\end{minipage}
\caption{State transitions for $\mathtt{CCPHASE_{011}}$ across all four pulses for all four combinations of $\mathtt{iSWAP_{02/20}}$; the forbidden combination ($\mathtt{iSWAP_{02}}$, $\mathtt{iSWAP_{20}}$) is shown in red with its erroneous output state. The evolution of each state is shown at the time steps depicted in the circuit to the left.}
\label{tab:allowed_combos}
\end{table*}
 \begin{figure}
     \centering
     \includegraphics[scale=0.21]{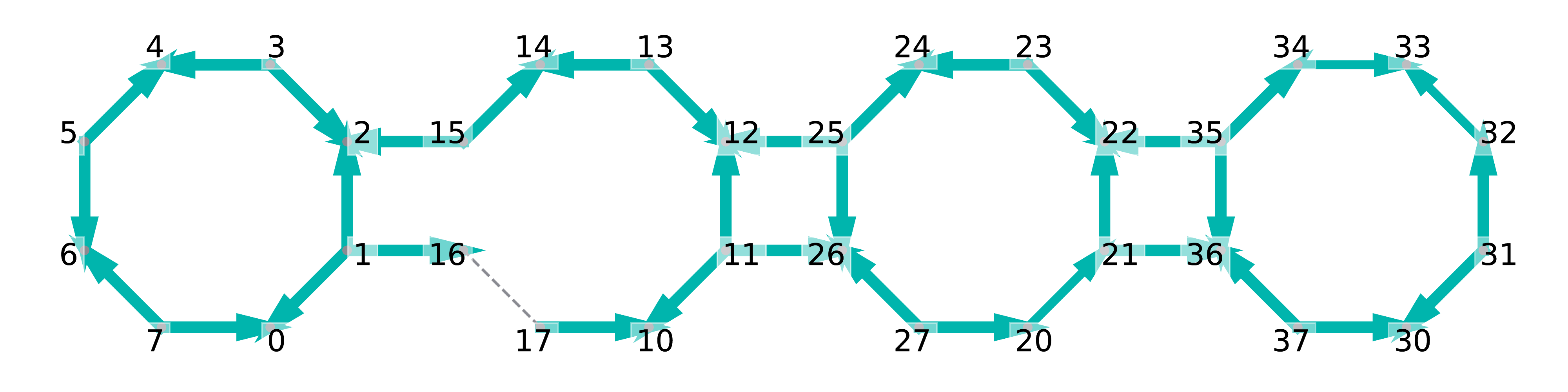}
     \caption{Interaction types for deployed $\mathtt{CPHASE}$ gates on Aspen-9, as of publication. Each arrow indicates the transmon that transitions to the second excited state $|2\rangle$.}
     \label{fig:gate_type}
 \end{figure}
Calibration of the qutrit-based $\mathtt{CCPHASE_{011}}$ gate over the entirety of our 32-qubit Aspen architecture requires careful mapping of which qubits swap to and from the second excited state. There are two interaction types for parametrically-activated $\mathtt{CPHASE}$ gates -- $\mathtt{iSWAP_{02/20}}$ -- possible on every edge of the Aspen platform; however, only a single type is calibrated for general use. The specific gate type chosen for each edge depends on how closely the gate frequency lies to other gates for a given AC flux-modulated qubit.  Figure~\ref{fig:gate_type} depicts the mapping of interaction types for deployed $\mathtt{CPHASE}$ gates on Aspen-9 - gate calibration is possible on any linear chain of three qubits in which the central qubit is not promoted to or from the second excited state in both of the constituent $\mathtt{CPHASE}$ interactions. We have described the specific sequence of $\mathtt{iSWAP_{20}} + \mathtt{iSWAP_{02}}$ on the sublattice of qubits (10, 11, 12), however, is possible to realize $\mathtt{CCPHASE_{011}}$ with other combinations of $\mathtt{iSWAP_{02}}$ and $\mathtt{iSWAP_{20}}$.

Note that, of the four possible choices, the combination ($\mathtt{iSWAP_{02}}$, $\mathtt{iSWAP_{20}}$) (where both operations rely on carrying $q_1$ into the second excited state) does not yield the intended gate unitary. \Cref{tab:allowed_combos} examines a subset of possible input states and illustrates how the disallowed combination - $\mathtt{iSWAP_{02}}$ + $\mathtt{iSWAP_{20}}$ - produces a unitary with an unwanted phase. In the case where $\theta = \pi$, the overall SU(8) unitary is equivalent to a circuit involving of two CZ gates and single-qubit rotations.

\bibliographystyle{apsrev}
\bibliography{combo_ref}

\end{document}